\RequirePackage{fix-cm}
\documentclass[twocolumn,epjc3]{svjour3}          
\smartqed  
\usepackage{graphicx}
\usepackage{enumitem}
\usepackage{mathtools}
\usepackage{amsmath}
\usepackage{xcolor}

\begin{document}

\title{Electron spectrometry with SDDs: a GEANT4 based method for detector
response reconstruction}


\author{Matteo Biassoni     \thanksref{infnmib}                 \and
        Matteo Gugiatti     \thanksref{polimi,infnmi}           \and
        Silvia Capelli      \thanksref{unimib,infnmib}          \and
        Marco Carminati     \thanksref{polimi,infnmi}           \and
        Oliviero Cremonesi  \thanksref{infnmib}                 \and
        Carlo Fiorini       \thanksref{polimi,infnmi}           \and
        Peter Lechner       \thanksref{hll}                     \and
        Susanne Mertens     \thanksref{mpi,tum}                 \and
        Lorenzo Pagnanini   \thanksref{unimib,infnmib}          \and
        Maura Pavan         \thanksref{unimib,infnmib}          \and
        Stefano Pozzi       \thanksref{unimib,infnmib}\thanks{Corresponding author, \email{stefano.pozzi@mib.infn.it}}
}


\institute{INFN - Sezione di Milano-Bicocca, 20126 Milano, Italy \label{infnmib}
            \and
            Dipartimento di Elettronica, Informazione e Bioingegneria, Politecnico di Milano, 20133 Milano, Italy \label{polimi}
            \and
            INFN - Sezione di Milano, 20133 Milano, Italy \label{infnmi}
            \and
            Dipartimento di Fisica G. Occhialini, Universit\`a degli Studi di Milano-Bicocca, 20126 Milano, Italy \label{unimib}
            \and
            Halbleiterlabor of the Max-Plank Society, 81739 Munchen, Germany \label{hll}
            \and
            Max Plank Institute fur Physics, 80805 Munchen, Germany \label{mpi}
            \and
            Technische Universitat Munchen, 80333 Munchen, Germany \label{tum}
}

\date{Received: date / Accepted: date}

\maketitle

\begin{abstract}
Electron spectrometry is traditionally challenging due to the difficulty of correctly reconstructing the original energy of the detected electrons. Silicon Drift Detectors, extensively used for X-ray spectrometry, are a promising technology for the precise measurement of electrons energy. The ability to correctly model the detector entrance window response to the energy deposited by electrons is a critical aspect of this application. We hereby describe a MonteCarlo-based approach to this problem, together with characterization and validation measurements performed with electron beams from a Scanning Electron Microscope.
\end{abstract}

\section{Introduction}
\label{intro}

SDDs (Silicon Drift Detectors) are flexible solid state devices characterised by a fast timing and an excellent energy resolution (close to the Fano limit in silicon). Thanks to these features and to a well-established technology, SDDs have revolutionized X-ray spectrometry in the last 40 years \cite{XrayWithSDD,Guazzoni}. 
Recently, their application in the field of beta spectroscopy was proposed \cite{tristan1,tristan2,tristan3,spettribeta}. As always when electrons are concerned, the challenge is to be able to infer with the required accuracy the energy of the impinging electron. Back-scattering, bremmstrahlung and dead layers are among the worrisome effects that can give rise to dramatic systematic errors such as sizable distortions of the energy spectra or biased determination of the real the source rate. The task is particularly tricky because an accurate reconstruction of the detector response function implies the use of a very well known source, devoid of any distortion in the emitted electron energy spectrum. In this paper, we discuss a technique based on the use of monochromatic electron beams as electron source whereas a GEANT4 simulation is used to model electron behaviour.

\section{Experimental setup} \label{sec:setup}
The hereby presented study has been performed with a single-pixel SDD used as detector, illuminated by a mono-energetic electron beam generated by a Scanning Electron Microscope (SEM), as well as an $^{55}$Fe X-ray source. A very detailed description of the setup used in this study and a full report on the results of the measurements at different beam energies and incident angles can be found in \cite{polipaper}. The process of optimization of the detector working point for X-rays and electrons is also described.

\subsection{SDD assembly and read-out} \label{subsec:sdd}
\label{sec:SDD}
\begin{figure}
  \includegraphics[width=\columnwidth]{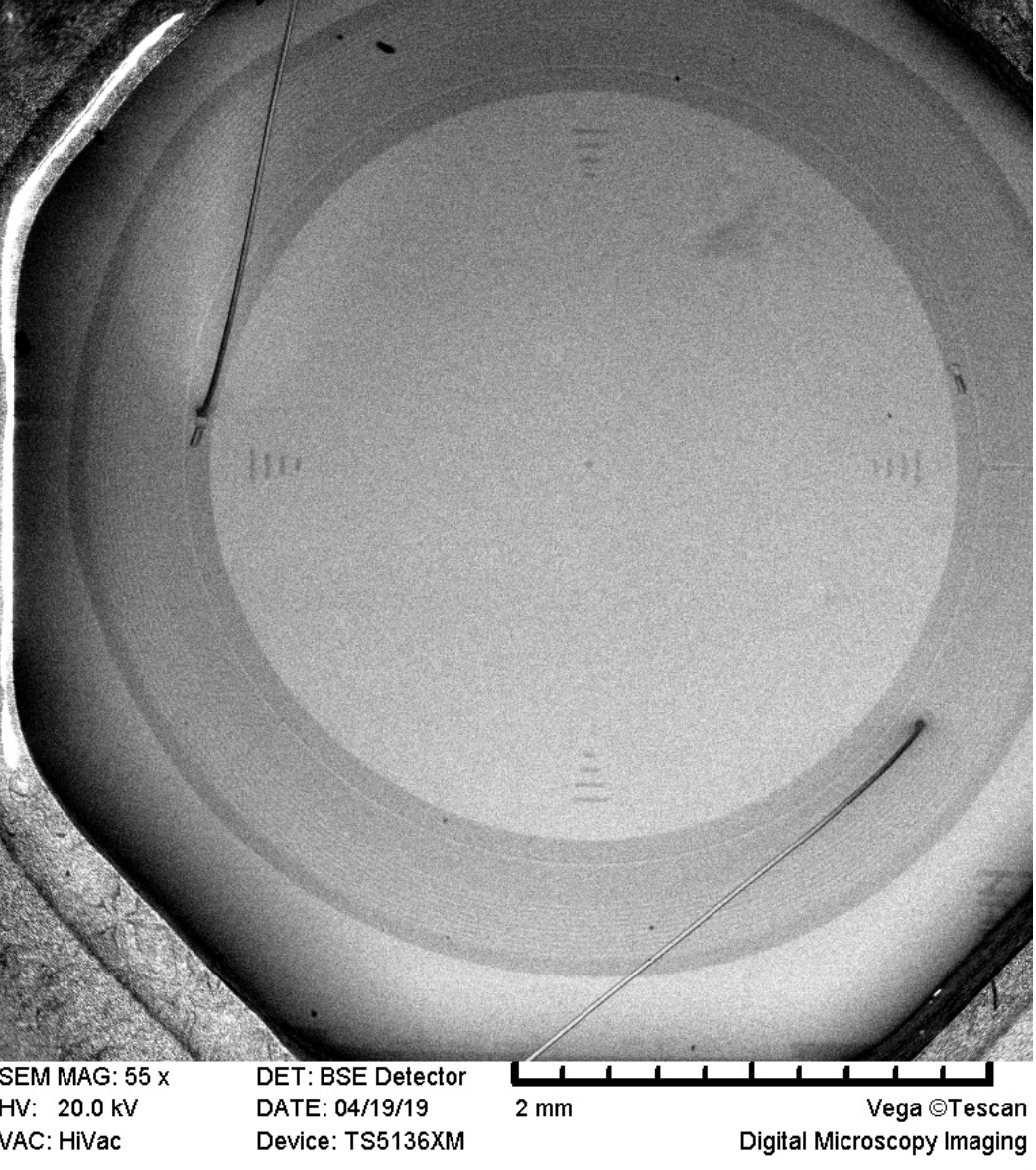}
\caption{SDD entrance window with the polarization contact and reference aluminum structures. The anode and integrated transistor sit below the central dot. The picture was realised with the SEM in imaging mode.}
\label{fig:SDD front}
\end{figure}

The SDD, fabricated by HLL-MPP,  has a circular design with 1600~$\mu$m radius active area (Figure~\ref{fig:SDD front}). The electric field is generated by polarizing a set of circular electrodes on the back of the device, while the front surface (where the radiation to be measured is supposed to hit the detector) is an equi-potential contact realized with a very shallow (of the order of tens of nanometers) implantation covered with a 22~nm thick silicon oxide passivated layer.
The entrance window has no metallic deposition in order to minimize the amount of passive material that the radiation must penetrate, but some reference aluminum structures identify the center of the device and the cardinal points, $\sim$100~$\mu$m away from the edge of the active area.
At the very center of the circular active area, where the collection anode is situated, an integrated transistor acts as the first stage of the charge integration and signal amplification chain.

The SDD is glued to an integrated circuit board hosting the readout electronics (front-end) where the anode signal is collected, the detector and transistor biases are generated, an ASIC is used as pre-amplifier \cite{Ettore} and the signal is routed towards a DAQ system. 
The DAQ consists of an XGLab Dante DPP (digital pulse processing unit) based on a 125 MHz, 16 bit digitizer. The programmable DPP applies a digital flat-top (trapezoidal) filter to the digitized waveform, and fills a histogram with the amplitude of the events.

The SDD and detector boards are covered with an aluminum protective case that also acts as the holder for the $^{55}
$Fe source. A hole in the cover allows the electron beam to hit the detector surface with a wide range of incidence angles. 

\subsection{SEM and electron beam} \label{subsec:sem}

\begin{figure}
  \includegraphics[width=\columnwidth]{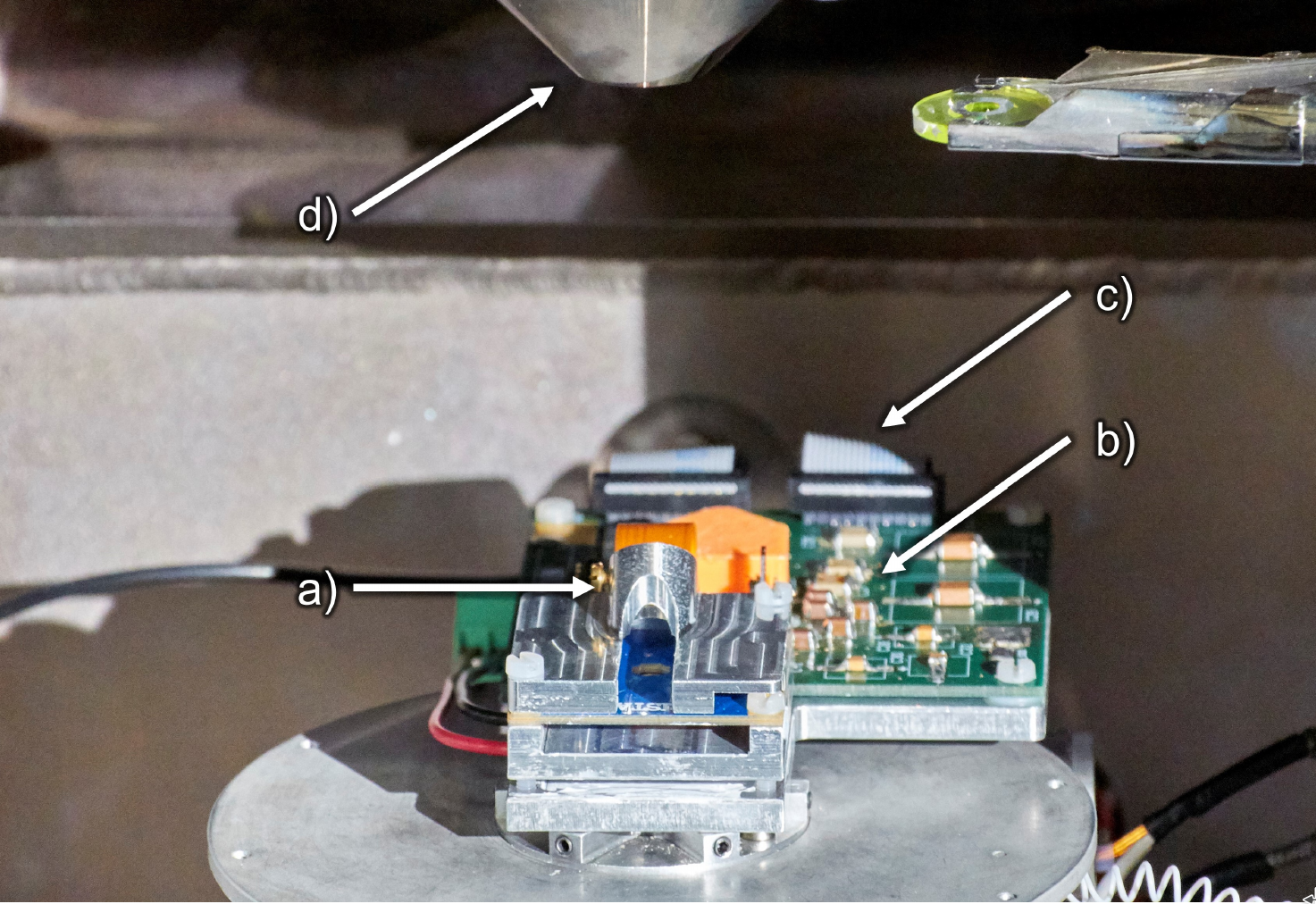}
\caption{Experimental setup with the SDD assembly on the SEM (\emph{Tescan VEGA TS 5136XM}) stage, inside the vacuum chamber. The setup consists of: (a) detector board with single-pixel SDD, (b) preamplifier board, (c) routing of the signal and biases towards the vacuum feed-through, (d) electron beam  source.}
\label{fig:setup}
\end{figure}

The electron beam used to characterize the SDD response function is generated by a \emph{Tescan VEGA TS 5136XM} SEM. The detector and front-end electronics are placed on the sample stage of the SEM (see Figure~\ref{fig:setup}), and connected via vacuum-compatible feed-throughs to the bias generators and DPP, placed outside the SEM vacuum chamber. The SEM stage can be moved with 1~$\mu$m precision along the x, y and z axis, rotated and tilted. The possibility of tilting the detector with respect to the electron beam axis is crucial for this study: the combined analysis of the energy spectra acquired in a wide range of incidence angles proves very sensitive to the profile of the entrance window energy-to-signal conversion. The larger is the beam angle w.r.t. the surface normal, the longer the effective thickness of the insensitive (or partially sensitive) part of the detector that the electrons must penetrate.

The beam emittance, as well as the uncertainty on its energy, are negligible compared to the size and energy resolution of the detector. The beam is therefore considered perfectly collimated and mono-energetic in the following analysis.

During the measurements, the current of the SEM beam has been reduced well below the minimum settings by reducing the heating current of the electrons source. In this way the event rate on the detector was low enough (in the ten kHz range) to make pile-up completely negligible.

\subsection{$^{55}$Fe X-ray source}
A real-time monitor of the energy scale and energy resolution of the SDD is performed thanks to the presence of the $^{55}$Fe source that contributes for about one-tenth of the overall counting rate measured by the detector ($\sim$3~kHz). The Fe source illuminates uniformly the surface of the SDD and the surrounding material.
As a consequence, in all the spectra the main full energy peaks (5.9 and 6.5 keV) are visible. In order to extend the comparison of the (real and simulated) spectra below $\sim$6 keV, a properly normalized spectrum of the $^{55}$Fe alone is added to the simulated electron spectrum before the calculation of the $\chi^2$ (Section~\ref{subsec:fit}).

\section{Detector response model} \label{sec:model}
One of the most tricky aspects of SDD characterization is the reconstruction of their energy-response function, $R(E;E_0,\alpha)$, that is defined as the probability of measuring an energy $E$ when the incoming electron energy is $E_0$ and the incident angle is $\alpha$. The main structures in this function are:
\begin{enumerate}[label=\arabic*)]
    \item a ``full energy" peak, with a FWHM that measures the energy resolution of the device. The position of this peak is slightly below the electron beam energy, due to a DC polarization of the entrance window of -110 V and to the effect of a partial signal generation efficiency at the surface (Section~\ref{subsec:QE});
    \item a left-side tail due to an incomplete collection of the charges produced by particle interaction;
    \item a roughly flat and featureless continuum at low energy due to a energy losses outside the detector volume, both from bremsstrahlung X-ray emission and electron back-scattering;
    \item silicon X-ray escape peaks generated when a silicon X-ray produced by the electron interaction leaves the detector.
\end{enumerate}

The structures described by 1), 3) and 4) are a consequence of the mechanism of interaction of the electrons with matter, and would be present also with an ideal detector with the ability of convert all the deposited energy into a readable signal. Their relative magnitude is obtained with the GEANT4 simulation described in Section~\ref{subsec:MC}.

The width of the full energy peak 1) is dominated by the combination of electron-hole pairs statistics and electronic noise. It is reproduced by convolving the output of the simulation with a gaussian function whose width is defined by
\begin{equation}
    \sigma(E) = \sigma_{\mathrm{noise}} \oplus \sigma_{\mathrm{e-h}} \sqrt{E}
\end{equation}
\noindent where $\sigma_{\mathrm{noise}}$ and $\sigma_{\mathrm{e-h}}$ are obtained by the analysis of X-ray calibration data.
    
The detector entrance window is the major source of the effect 2). Its treatment is described in Section~\ref{subsec:QE}.
An alternative approach to account for the effect of the entrance window would be an event-by-event simulation of the full device that describes charge production, drift and collection. This approach would be too computationally demanding and complex to be implemented in a real-life application, where an effective model is much more efficient when dealing with a large number of events of a typical application \cite{tristan1,tristan2,spettribeta}. 

In the case of X-rays, the entrance window has a mild effect on $R(E)$: most of the photons interact in the very bulk of the detector where charge collection is fully efficient. The $R(E)$ still features a gaussian peak, corresponding to the X-ray nominal energy, and the low energy continuum, but the effect of the entrance window is negligible.
In the case of electrons, the effect of the entrance window is more dramatic because no electron can reach the fully active detector volume without experiencing an energy loss in the shallower layers. The peak is shifted toward lower energies (by an amount proportional to the minimum amount of energy that the electron can loose in the inactive volume when crossing it perpendicularly with minimal deviations) and deformed. The low energy continuum is also more pronounced, mostly because of the high probability of electron back-scattering.
Moreover, by changing the incidence angle of the electron beam with the SDD surface, the relative importance of the different effects can be changed. The further from normal incidence, the larger the amount of energy lost in the inactive and partially active parts of the entrance window, and the larger the probability of backscattering.
For this reason an electron beam is an optimal solution for this characterisation.

\subsection{MonteCarlo simulation of beam interaction} \label{subsec:MC}
We use GEANT4 \cite{geant4} to simulate the interaction of an electron of energy $E_0$ entering the SDD with an angle $\alpha$. The simulation predicts the punctual energy depositions $E_i(x,y,z)$ along the electron track inside the material. At each step the different energy-depositing processes (continuous energy loss, nuclear diffusion) as well as the production and propagation of secondary radiation (bremsstrahlung, photo-electric effect, Rayleigh and Thomson scattering, X-ray escape) are considered.
The detector is modeled as a homogeneous layer of silicon. Relevant parts of the setup located near the detector active volume (PCB, detector holders, etc.) are also included in the simulation in order to account for the contribution to the signal of escaping radiation re-entering the detector after interacting with the surrounding material. For each value of $E_0$ and $\alpha$ a large number of electrons (events) is simulated, and for each event the coordinates of each energy deposition are recorded. This information is later weighted for the Q.E. (section~\ref{subsec:QE}) to extract the expected signal magnitude.

As the energy of the electrons is relatively low (few of keV) and the required spatial accuracy of the simulation very high (the SDD Q.E. is expected to show relevant variations on the scale of tens of nanometers), the usual statistical treatment of the energy depositions can lack the required accuracy and some non-standard settings of GEANT4 (SingleScattering and G4StepLimiter feature) have been tested looking for the configuration that better reproduces the data.
The choice of the physics processes to be included in the simulation and their parametrisation has been validated by our group in various experimental conditions with a variety of detectors, from the conventional scintillators and semiconductor detectors to thermal detectors~\cite{CUORE_BB,CUORE0_BM,CUPID0_2nu}.

\subsection{Energy-to-signal conversion efficiency} \label{subsec:QE}

Once the information about the energy deposited by each event at different coordinates inside the SDD is generated by the GEANT4 simulation and recorded, an analytical model ($f_{\mathrm{QE}}(z;\theta)$) of the energy-to-signal conversion (Q.E.) is applied to the electron sample in order to build the expected spectrum of recorded amplitudes.
The recorded amplitude is defined as a weighted sum of the energy depositions:
\begin{equation}
    E=\sum E_i(x,y,z) \cdot f_{\mathrm{QE}}(z;\theta)
\end{equation}
The Q.E. model $f(z;\theta)$, acting as a weight function, is assumed to depend only on the $z$ coordinate (but the study could, in principle, be extended to a higher dimensional model), i.e. the penetration depth. $f$ also depends on a set of parameters, $\theta$.

The general shape of the Q.E. reflects the technology used to build the SDD entrance window: an oxide layer exists on the device surface (any electron-hole pair produced here contributes to the signal formation with a constant probability, $p_0$, that is expected to be zero but is still left as a free parameter in the optimization process), followed by a volume where the charge collection is at work but not fully efficient. For this particular device the thickness of the oxide dead layer, $t$,
is expected to be $\sim$22 nm.
Increasing $z$ beyond the boundary between the oxide and the active silicon, a volume where the implantation of the entrance window contact occurred is encountered. Here the Q.E. is assumed to start from a value $p_1$ and gradually approaches a unitary value with a regular behaviour that we decided to model as an exponential with scale parameter $\lambda$\cite{SDDwindow,SDDwindow2}. Once the free parameters of the model, $\theta = \{t,p_0,p_1,\lambda\}$, are defined, the Q.E. can be written as:

\begin{equation*}
    f_\mathrm{QE}(z;t,p_0,p_1,\lambda) = \begin{dcases*} 
        p_0 & $z < t$ \\ 
        1+(p_1-1)\exp{\left(-\frac{z-t}{\lambda}\right)} &  $z > t$
    \end{dcases*}
\end{equation*}

The aim of this study is to show that a set of optimized parameters, $\hat\theta$, can be obtained by comparing the simulated spectra with real data. The resulting response function can be used to predict the shape of the measured spectrum for any source of electrons illuminating the SDD. In this particular work, in order to significantly speed up the minimization process, we assume $p_0 = 0$ and $t = 22$~nm, as the oxide layer thickness is known from the design of the device and fabrication process. 

\begin{figure}
    \includegraphics[width=\columnwidth]{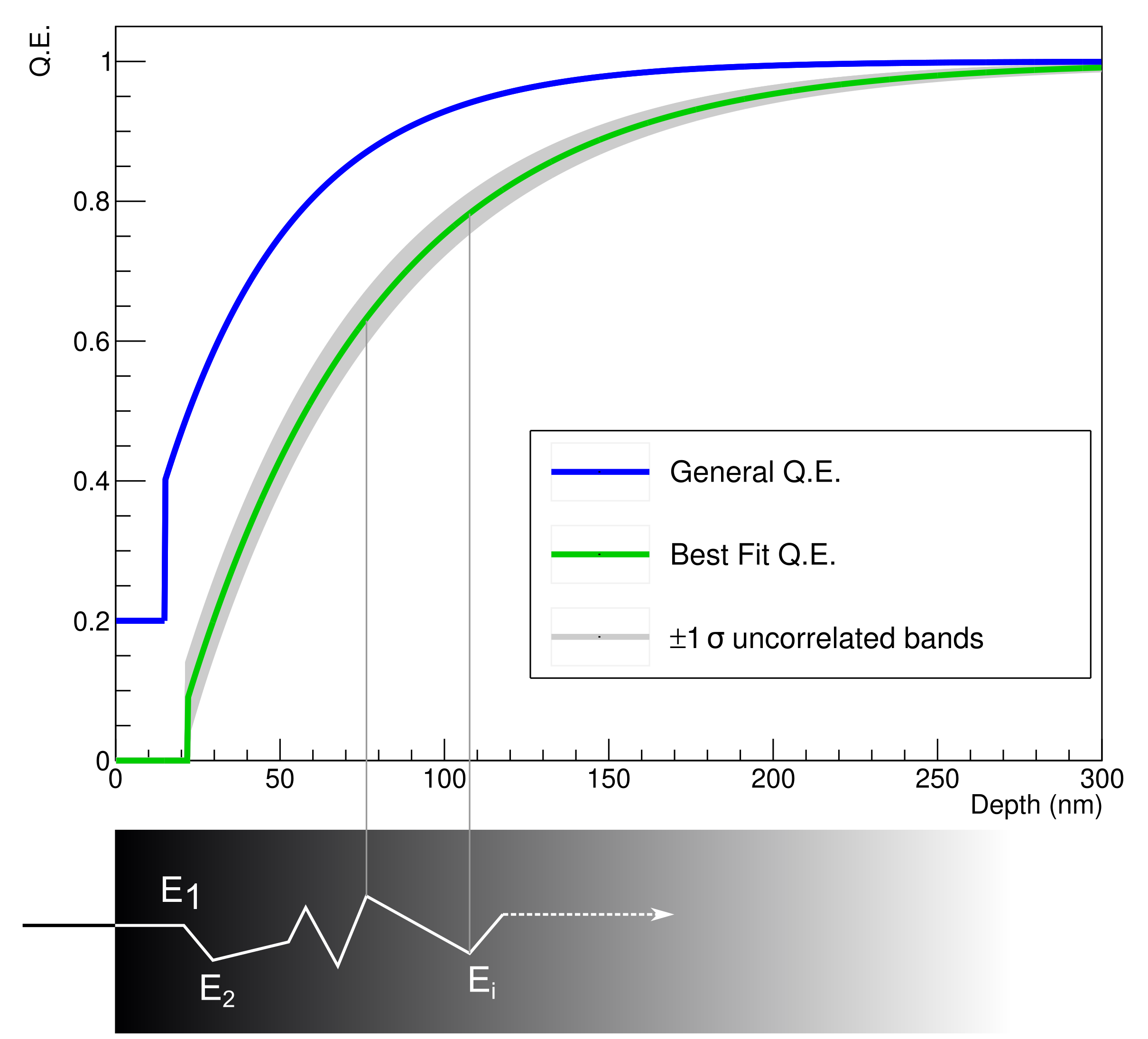}
        \caption{Energy-to-signal conversion efficiency (Q.E.) parametric model. In blue the most generic form of the Q.E., with a shallow layer of partial efficiency followed by an exponentially increasing region with free intercept. The green curve is the configuration eventually found to better fit the data, with gray bands representing the curves obtained by varying by $\pm1\sigma$ the free parameters in an uncorrelated way. In the lower panel a conceptual sketch of a recorded MonteCarlo event for a low energy electron. The total signal is evaluated as the energy E$_i$ deposited along each step, weighted for the Q.E. evaluated at the corresponding coordinates.}
        \label{fig:QE}
\end{figure}

\subsection{Fitting procedure and estimation of model parameters} \label{subsec:fit}

The estimation of the best parameters $\hat\theta$ is performed with a simultaneous least square fit of the simulations to the available data-sets (different energies and angles). The data used for this analysis have been collected with three beam energy settings (5, 10 and 20 keV) and incidence angles varied between 5° and 60° with a step of 5°. It's worth noting that the real energy $E_0$ of the beam is assumed to be 110 eV smaller than the energy setting of the SEM, due to the polarization of the entrance window contact.
For each beam condition (defined by energy $E_0$ and incidence angle $\alpha$) we run a high statistics (in order to make the statistical errors of the model subdominant) simulation recording the punctual energy depositions $E_i(x,y,z)$ occurring in the SDD. We then process each event in the simulation with the selected $f(z,\theta_j)$ model to obtain the simulated energy spectrum that we expect to measure if the Q.E. were described by the set of parameters $\theta_j$
that we compare with the experimental one.
We sample the $\theta$ parameter space and for each point ($\theta_j$) we perform a simultaneous comparison of the real and simulated spectra for all available $E_0$ and $\alpha$. From this comparison we extract a $\chi^2$ value for each point $j$ in the $\theta$ space as:
\begin{equation}\label{eq:chi2global}
    \chi^2_j = \sum_{E_0,\alpha} \chi^2_{E_0,\alpha;j}.
\end{equation}

Each term in the sum in Equation~\ref{eq:chi2global} is defined as

\begin{equation}
    \chi^2_{E_0,\alpha;j} = \sum_n \frac{(D_n(E_0,\alpha) - S_n(E_0,\alpha;\theta_j))^2}{D_n(E_0,\alpha)},
\end{equation}

where $D_n(E_0,\alpha)$ is the number of events in the $n$-th bin of the data with beam energy $E_0$ and incidence angle $\alpha$. $S_n(E_0,\alpha;\theta_j)$ is the content of the $n$-th bin in the corresponding simulation, processed with a the $j$-th set of model parameters $\theta_j$. We exclude the bins corresponding to energies smaller then 3 keV from the $\chi^2$ calculation as threshold effects, as well as random triggers of noise fluctuations introduce distortion in this energy region that we don't model in our simulation. Similarly, we consider only energies up to 200 eV (1 keV) above the beam energy for the 5 keV (10 and 20 keV) data-sets. We don't expect any relevant information above these thresholds.

\begin{figure}
  \includegraphics[width=\columnwidth]{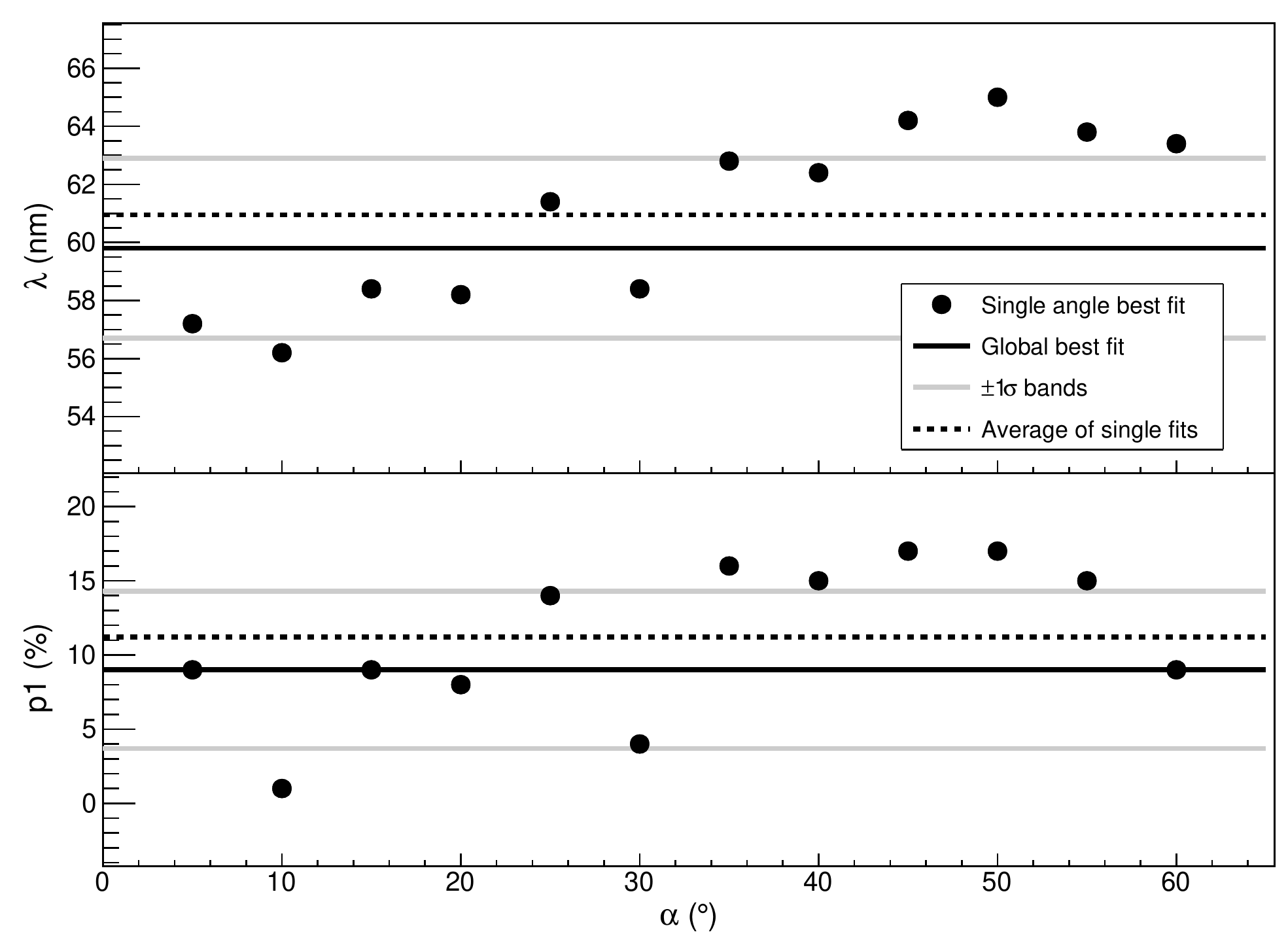}
\caption{Values of the two free parameters $\lambda$ and $p_1$ obtained from the fits of the single angle ($\alpha$) data-sets. For reference, the average of the points (dashed), as well as the global fit result (solid black) and the associated errors (solid grey) are reported. The result of the fit is stable with respect to the choice of the subset of analyzed data, and the spread of the results is therefore used as a measure of the error associated to the global fit outcome.}
\label{fig:sys}       
\end{figure}

The minimum value among the $\chi^2_j$ corresponds to the best set of parameters $\tilde\theta$. We extract statistical errors on these parameters by interpolating the region around the minimum and marginalizing.
In order to account for possible biases in the estimation of the parameters deriving from the fact the model that we used might be an over-simplification of the real response, we repeat the fitting procedure on reduced data-sets corresponding to 
single angles (summing over the different energies). We therefore obtain a total of 12 best fit values (Figure~\ref{fig:sys}), and we report their RMS as an estimation of the systematic error associated to the best fit from the simultaneous fit.
The resulting values for the free parameters are the following:
\begin{equation*}
    \lambda = (59.8 \pm 3.1) \mathrm{~nm}
\end{equation*}
\begin{equation*}
       p_1 = (9 \pm 5)\%
\end{equation*}
As expected given the large amount of statistics in the single spectra, the statistical error is negligible and the uncertainty on the parameters is completely dominated by the systematics contribution, accounting for the small deviations of the model from the data in some regions of the spectrum.

\begin{figure}
  \includegraphics[width=\columnwidth]{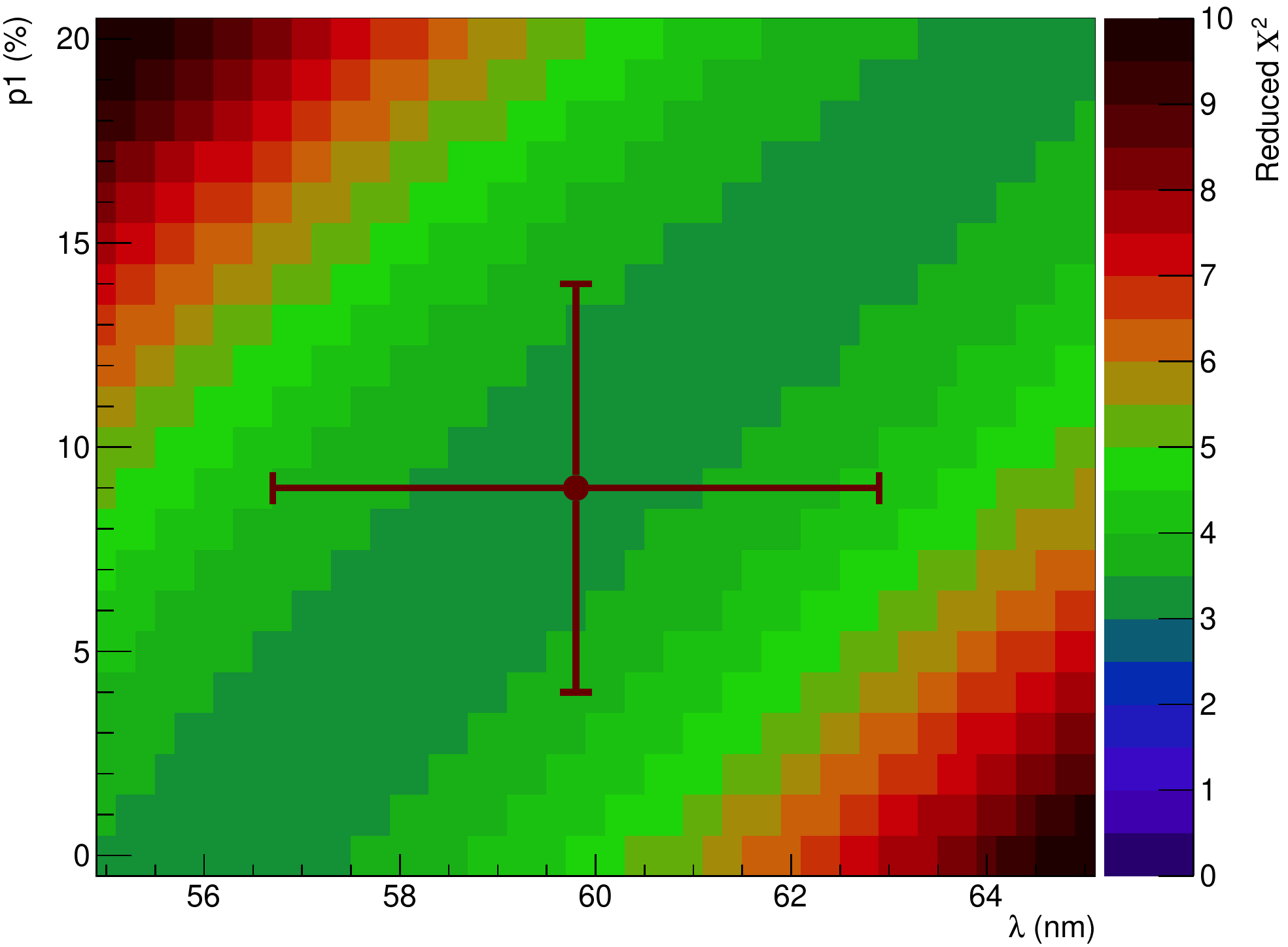}
\caption{Projection of the reduced $\chi^2$ in the $(\lambda,p_1)$ plane. Minimum and correlations between parameters can be extracted from these projections. The error bars are dominated by systematic effects that have been evaluated as the RMS of the results obtained by repeating the fit on the 12 independent subsets of data corresponding to single incident angles.}
\label{fig:chi2proj}       
\end{figure}

In Figure~\ref{fig:chi2proj} a bi-dimensional projection of the reduced $\chi^2$ in the $(\lambda,p_1)$ plane is reported for reference. Together with the minimum, corresponding to $\tilde\chi^2 = 3.3$, correlations between the parameters can be extracted with this method.

\begin{figure}
  \includegraphics[width=\columnwidth]{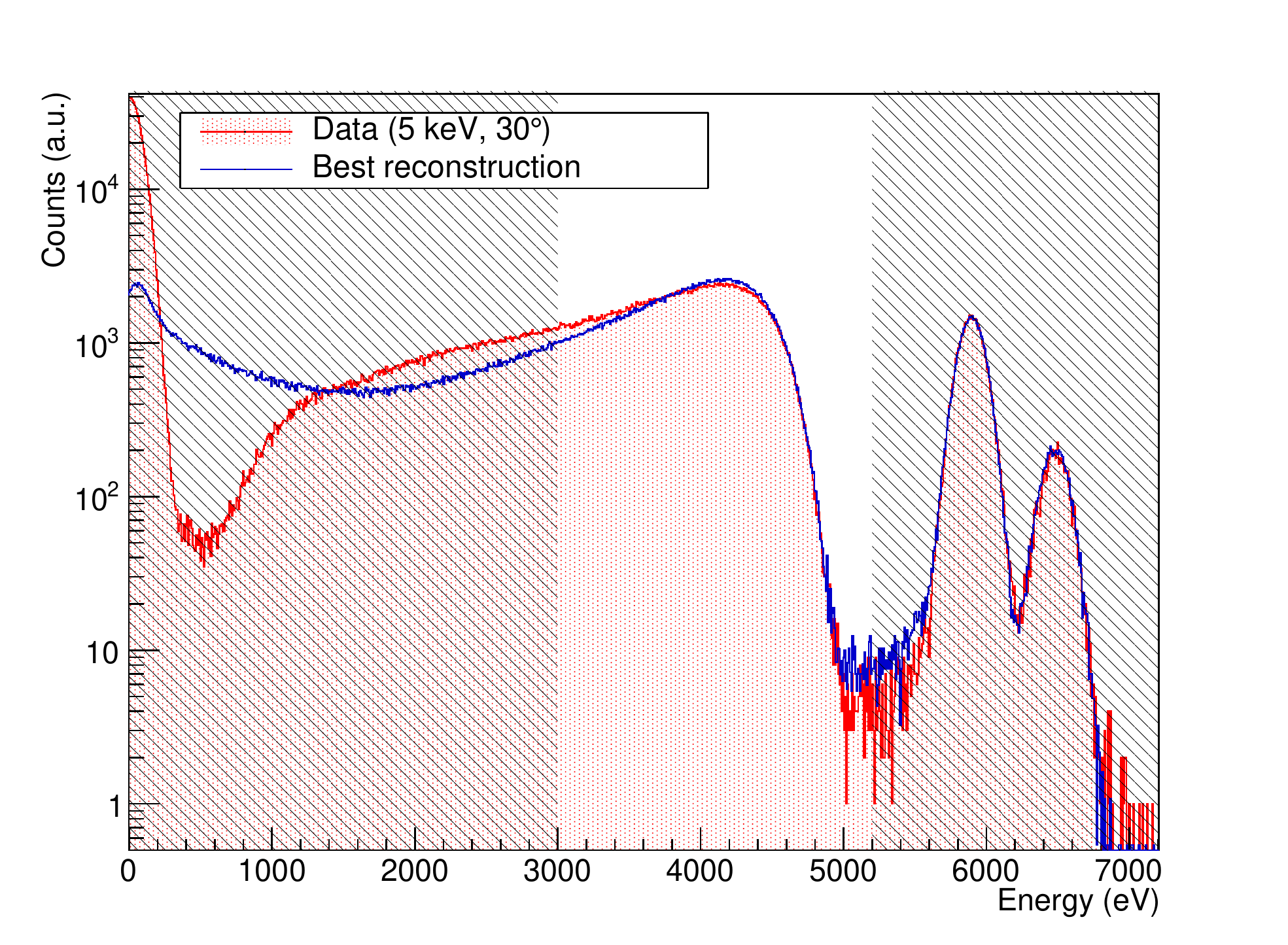}
\caption{Data (red) and MC reconstruction (blue) of the response function to 5 keV electrons hitting the surface of the SDD with a 30° angle. The reconstruction is obtained with the best fit response function from the global analysis of the different energies and incidence angles. The $\chi^2$ is calculated only in the non-shaded regions. The region below 3 keV is always excluded because threshold and random trigger (noise) effects are not modelled in our simulation, while the upper limit is slightly above the beam energy.}
\label{fig:bestfit1}       
\end{figure}

\begin{figure}
  \includegraphics[width=\columnwidth]{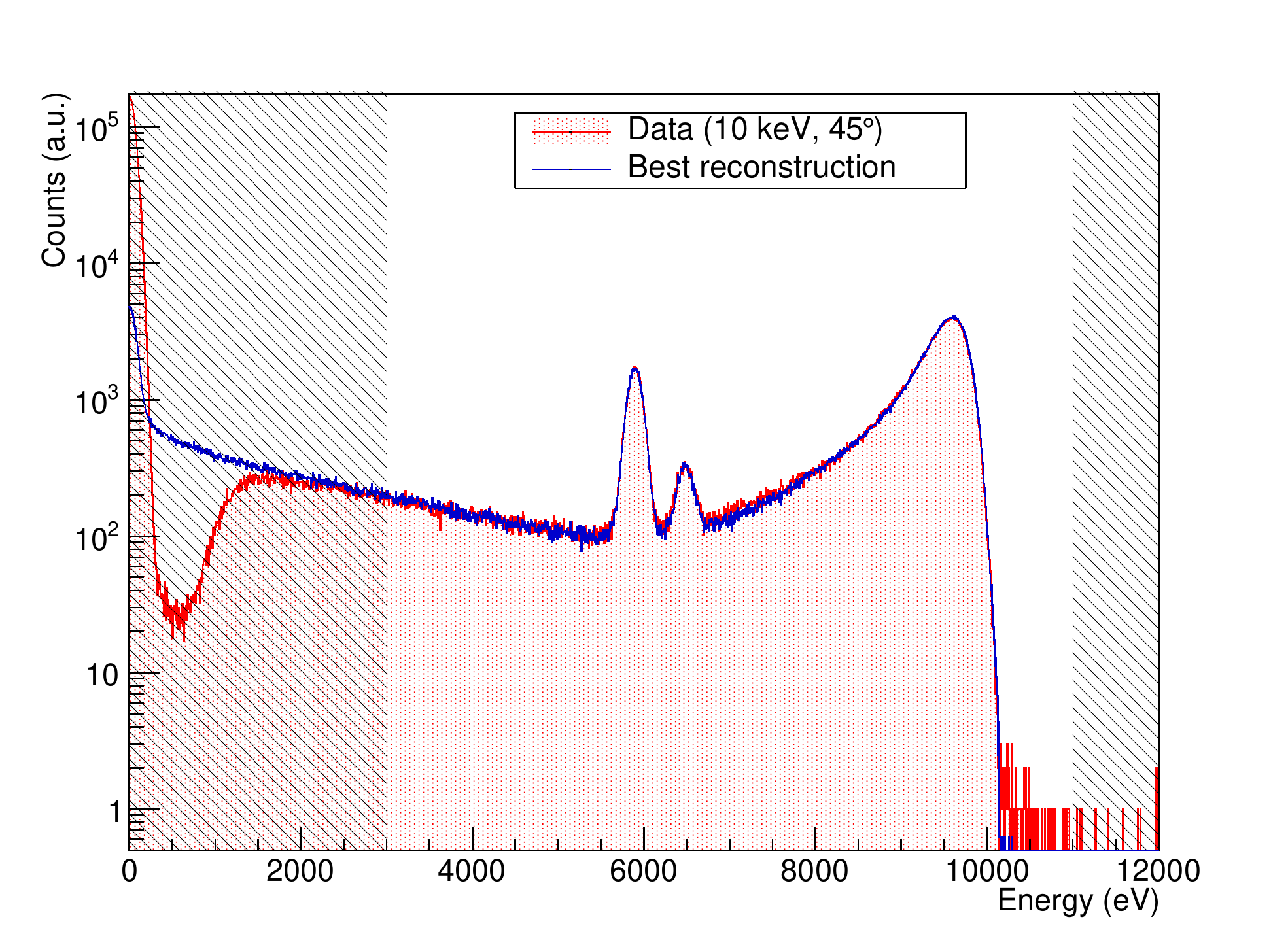}
\caption{Data (red) and MC reconstruction (blue) of the response function to 10 keV electrons hitting the surface of the SDD with a 45° angle. The $\chi^2$ is calculated only in the non-shaded regions. The region below 3 keV is always excluded because threshold and random trigger (noise) effects are not modelled in our simulation, while the upper limit is slightly above the beam energy.}
\label{fig:bestfit2}       
\end{figure}

\begin{figure}
  \includegraphics[width=\columnwidth]{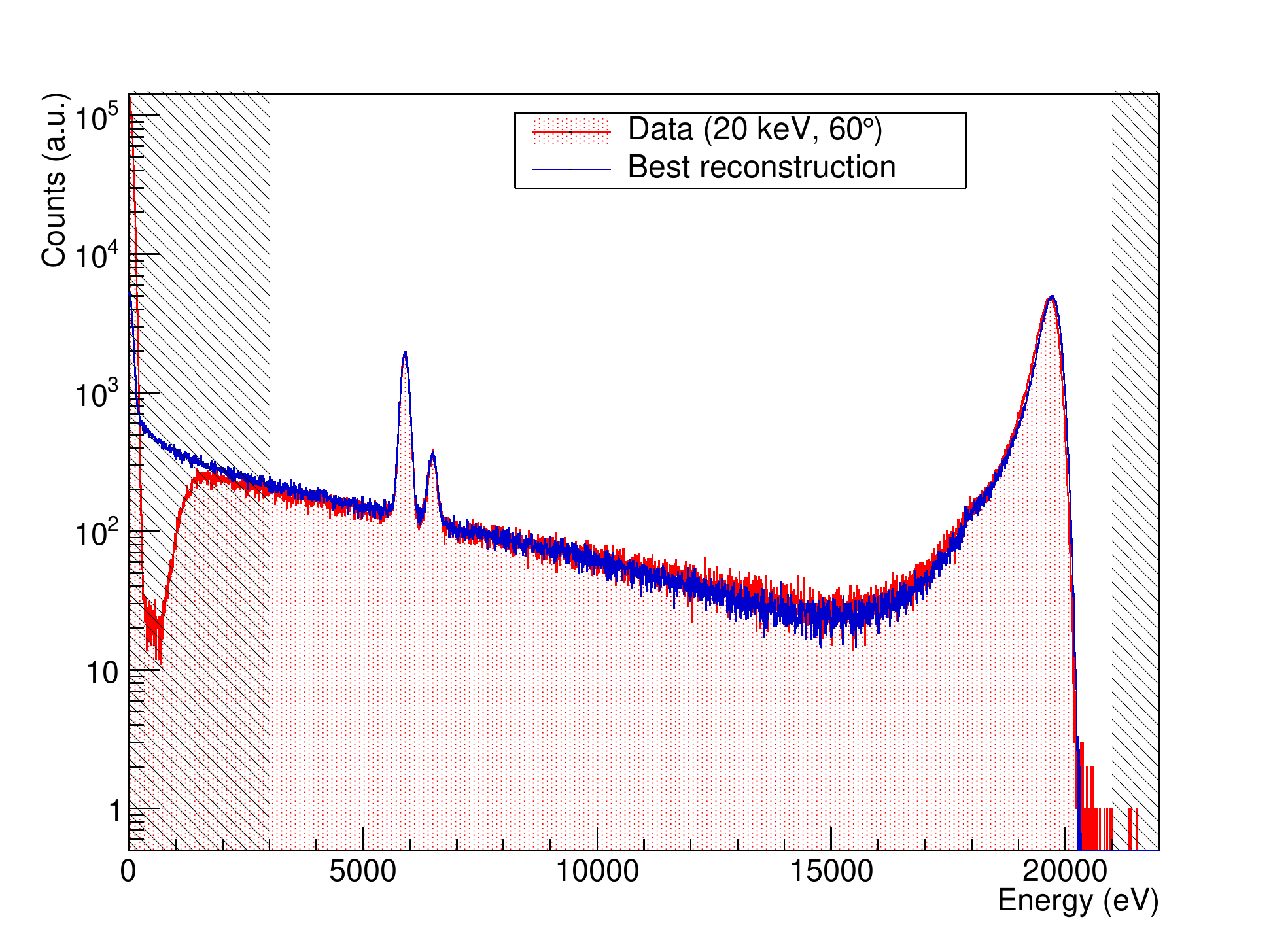}
\caption{Data (red) and MC reconstruction (blue) of the response function to 20 keV electrons hitting the surface of the SDD with a 60° angle. The $\chi^2$ is calculated only in the non-shaded regions. The region below 3 keV is always excluded because threshold and random trigger (noise) effects are not modelled in our simulation, while the upper limit is slightly above the beam energy.}
\label{fig:bestfit3}       
\end{figure}

Figure~\ref{fig:bestfit1}, \ref{fig:bestfit2} and \ref{fig:bestfit3} show three examples of comparison between data (red) and the MonteCarlo reconstruction (blue) performed with the best set of parameters $\hat\theta$ for three of the $(E_0,\alpha)$ combinations. 
The global agreement between the data and the model is remarkably good in the wide range of energies and angles analysed, with the maximum deviation over all the 36 spectra being of the order of $10\%$ of the peak intensity over few (out of many thousands) energy bins.

\section{Conclusions}
This work presents a method to develop a model of the response of an SDD to keV electrons. The response function is built by comparing GEANT4 simulations, processed with an analytical model of the Q.E. of the SDD entrance window with data acquired when illuminating a device with mono-energetic electron beams and X-rays. A precise knowledge of the device response function is crucial for any application of SDDs requiring systematics-free spectroscopy of electrons: from fundamental physics searches\cite{tristan3,spettribeta,Suhonen} to radio-protection and metrology where SDD could represent a cheap and compact alternative to classical instruments and methods\cite{radioprotezione,90Sr}.

\section*{Acknowledgments}
The measurements analyzed in this work has been acquired in the Electron Microscopy laboratory of the Department of Material Science of the University of Milano-Bicocca. We thank Prof. M. Acciarri and Dr. P. Gentile for their precious help and competence.

%




\end{document}